\documentclass{spi-b2474}

\usepackage{float}
\usepackage{xcolor}

\def\jnl@style{\it}
\def\aaref@jnl#1{{\jnl@style#1}}

\def\aaref@jnl#1{{\jnl@style#1}}

\def\aj{\aaref@jnl{AJ}}                   
\def\araa{\aaref@jnl{ARA\&A}}             
\def\apj{\aaref@jnl{ApJ}}                 
\def\apjl{\aaref@jnl{ApJ}}                
\def\apjs{\aaref@jnl{ApJS}}               
\def\ao{\aaref@jnl{Appl.~Opt.}}           
\def\apss{\aaref@jnl{Ap\&SS}}             
\def\aap{\aaref@jnl{A\&A}}                
\def\aapr{\aaref@jnl{A\&A~Rev.}}          
\def\aaps{\aaref@jnl{A\&AS}}              
\def\azh{\aaref@jnl{AZh}}                 
\def\baas{\aaref@jnl{BAAS}}               
\def\jrasc{\aaref@jnl{JRASC}}             
\def\memras{\aaref@jnl{MmRAS}}            
\def\mnras{\aaref@jnl{MNRAS}}             
\def\pra{\aaref@jnl{Phys.~Rev.~A}}        
\def\prb{\aaref@jnl{Phys.~Rev.~B}}        
\def\prc{\aaref@jnl{Phys.~Rev.~C}}        
\def\prd{\aaref@jnl{Phys.~Rev.~D}}        
\def\pre{\aaref@jnl{Phys.~Rev.~E}}        
\def\prl{\aaref@jnl{Phys.~Rev.~Lett.}}    
\def\pasp{\aaref@jnl{PASP}}               
\def\pasj{\aaref@jnl{PASJ}}               
\def\qjras{\aaref@jnl{QJRAS}}             
\def\skytel{\aaref@jnl{S\&T}}             
\def\solphys{\aaref@jnl{Sol.~Phys.}}      
\def\sovast{\aaref@jnl{Soviet~Ast.}}      
\def\ssr{\aaref@jnl{Space~Sci.~Rev.}}     
\def\zap{\aaref@jnl{ZAp}}                 
\def\nat{\aaref@jnl{Nature}}              
\def\iaucirc{\aaref@jnl{IAU~Circ.}}       
\def\aplett{\aaref@jnl{Astrophys.~Lett.}} 
\def\apspr{\aaref@jnl{Astrophys.~Space~Phys.~Res.}}
\def\bain{\aaref@jnl{Bull.~Astron.~Inst.~Netherlands}}
\def\fcp{\aaref@jnl{Fund.~Cosmic~Phys.}}  
\def\gca{\aaref@jnl{Geochim.~Cosmochim.~Acta}}   
\def\grl{\aaref@jnl{Geophys.~Res.~Lett.}} 
\def\jcp{\aaref@jnl{J.~Chem.~Phys.}}      
\def\jgr{\aaref@jnl{J.~Geophys.~Res.}}    
\def\jqsrt{\aaref@jnl{J.~Quant.~Spec.~Radiat.~Transf.}}
\def\memsai{\aaref@jnl{Mem.~Soc.~Astron.~Italiana}}
\def\nphysa{\aaref@jnl{Nucl.~Phys.~A}}   
\def\physrep{\aaref@jnl{Phys.~Rep.}}   
\def\physscr{\aaref@jnl{Phys.~Scr}}   
\def\planss{\aaref@jnl{Planet.~Space~Sci.}}   
\def\procspie{\aaref@jnl{Proc.~SPIE}}   

\makeatletter
\newcommand\listexamplename{List of Examples}
\newcommand\listofexamples{%
    \if@twocolumn
      \@restonecoltrue\onecolumn
    \else
      \@restonecolfalse
    \fi
    \chapter*{\listexamplename}%
      \@mkboth{\listexamplename}{\listexamplename}%
    \@starttoc{loe}%
    \if@restonecol\twocolumn\fi
    \cleardoublepage
    }
\newcommand*\l@example{\@dottedtocline{1}{0em}{3.6em}}
\makeatother

\newcounter{infobox}[chapter]
\renewcommand{\theinfobox}{\thechapter.\arabic{infobox}}
\renewenvironment{example}[1]{%
    \refstepcounter{infobox}
    \refstepcounter{example}
    \addcontentsline{loe}{example}{\protect\numberline{\hspace{1.5em}\theexample}\hspace{2.5em}#1}\par
    \begin{mdframed}[%
        frametitle={\large Example \theinfobox: \emph{#1}},
        backgroundcolor=gray!30,
        roundcorner=10pt,
        linecolor=black!60,
        outerlinewidth=1.5,
    ]}{\end{mdframed}}

\newtheorem{exa}{Example}
\newcounter{box}[chapter]
\renewcommand{\thebox}{\thechapter.\arabic{box}}

\newtheorem{exap}{Example}
\newcounter{Box}[chapter]
\renewcommand{\theBox}{\thechapter.\arabic{Box}}

\makeatletter
\newcommand{\ion}[2]{#1$\;$\textsc{\rmfamily\@roman{#2}}\relax}
\makeatother

\def\cleardoublepage{\newpage}

\makeatletter
\renewcommand\@biblabel[1]{}
\makeatother

\begin{document}














\begin{center}
{\Large Studying distant galaxies:\\
A Handbook of Methods and Analyses}\\
\end{center}
\begin{center}
F. Hammer$^1$, M. Puech$^1$, H. Flores$^1$, M. Rodrigues$^1$\\
\end{center}
\begin{center}
$^1$GEPI, Observatoire de Paris, CNRS, Univ. Paris Diderot, Place Jules Janssen, 92190 Meudon, France 
\end{center}
\begin{center}

ABSTRACT \\
\end{center}

\begin{center}
\begin{minipage}[t]{10cm}
Distant galaxies encapsulate the various stages of galaxy evolution and formation from over 95\% of the development of the universe. As early as twenty-five years ago, little was known about them, however since the first systematic survey was completed in the 1990s, increasing amounts of resources have been devoted to their discovery and research. This book summarises for the first time the numerous techniques used for observing, analysing, and understanding the evolution and formation of these distant galaxies. 

In this rapidly expanding research field, this text is an every-day companion handbook for graduate students and active researchers. It provides guidelines in sample selection, imaging, integrated spectroscopy and 3D spectroscopy, which help to avoid the numerous pitfalls of observational and analysis techniques in use in extragalactic astronomy. It also paves the way for establishing relations between fundamental properties of distant galaxies. At each step, the reader is assisted with numerous practical examples and ready-to-use methodology to help understand and analyse research.

{\it   Here is presented an excerpt of about 20\% of the book content. }
\end{minipage}
\end{center}

\chapter*{Introduction}

Galaxies are complex objects comprising hundreds of billions of stars, multiple
gas phases, and dust. Present-day galaxies are organized into a well-defined
sequence, the Hubble Sequence. Galaxy complexity naturally increases with
redshift, since distant galaxies are likely the subject of transformation
mechanisms. The study of distant galaxies is particularly enthralling since the
Cosmological Principle infers that they are causally linked to the present-day
ones.

There are considerable difficulties in studying distant galaxies. The more distant
they are, the more they bear witness to the \hbox{earliest} stages of galaxy formation,
but this is accompanied by faintness, smallness, and redshifted emissions, all
challenges for interpreting their observation. Theoretical interpretations require
the determination of stellar and gas mass, star formation, and heavy element
abundances, all of which are quantities obtained indirectly, and which are
afflicted by uncertainties and systematics both large and numerous. Galaxies are
also not homogeneously distributed in space, and their spectral energy
distributions are affected by redshift, stellar ages, abundances and dust, thus
complicating the rationale of their selection.

This handbook includes a state-of-the-art description of the sampling and
selection effects, based on observations with current instruments at large
telescopes and using complementary approaches including photometry, imagery, as
well as integrated and spatially-resolved spectroscopy. Such methods can be found
in a myriad of scientific papers or reviews, synthetized for the first time in a
single book. It aims to be a vital tool for teaching final-year undergraduate and
postgraduate courses on the topics of ``Galaxies'' and “Observational
Cosmology”. A number of illustrated examples will assist the reader in deriving
various galaxy physical quantities from observables. The book shows or proposes
numerous methods to derive physical quantities from observations, but also
expresses reasonable doubts about far-reaching conclusions made on minimal or
insufficient data. Masters and PhD students are encouraged to use it while
preparing research projects, including telescope time proposals or articles.
Academics may inquire about the numerous aspects surrounding their field of
research, which they can better evaluate.


The first chapter establishes how a galaxy sample can be gathered and the many associated
caveats which may hamper theoretical interpretations. It addresses the statistical
completeness issue, the redshift measurement, the luminosity function analysis, and
finally the possibility to causally link galaxy samples from different epochs. The second
chapter provides an overview of the basics of imagery and photometry and their
uncertainties, as well as the heightened complexity of morphological analyses. A method
to derive the latter analyses is proposed, which minimizes systematics while allowing to
derive stellar mass and star formation from the energy distribution. The third chapter
provides a comprehensive basis for integrated spectroscopy, showing how one can derive
the heavy element abundances of their gas phases and star formation rates (SFRs), as well
as determine the metal abundances from UV absorption lines and constrain the stellar
populations. In principle, velocity is the most directly determined parameter for distant
galaxies, and it is the role of Chapter 4 to introduce the considerable limitations
caused by galaxy faintness and smallness. It then includes a full description of existing
methods for reducing data cubes with a comprehensive treatment of sky residuals, and
proposes a method to derive then classify the dynamical state of galaxies, called
``morpho-kinematics''. Chapter 5 describes how to synthesize the numerous observations
provided for a single, distant galaxy, to determine its physical properties and allow its
modeling. It addresses the important question of stellar mass evaluation as well as that
of the gas in its various phases. It finally shows the subsequent strengths and
weaknesses of scaling relations between mass, velocity, abundance, and star formation.

\chapter{Samples and Selection Effects}
{\it Here it is an excerpt of Chapter 1.}
\section{Pre-selection of a sample prior to redshift measurement}
\label{C1_Sect1.2}
Cosmological variance plays an important role in studies of distant galaxies.
Variations of the galaxy density are expected at clustering scales of 5 Mpc to
10 Mpc, implying a minimal size of a deep pencil beam of $\sim$100 ${\rm arcmin}^{2}$.
Nevertheless, large scale structures like filaments are larger than that and some
well-studied fields (e.g., the \gls{GOODS}-South field and the scientific results derived
from them may be significantly affected, see (Gilli et al., 2003, Vanzella et al.,2005, Ravikumar et al., 2007).

Typical variations in luminosity density can be up to a factor 2 from one field to
another and large scale structures may have a considerably effect on the LF that one
derives (see Sec.~\ref{C1_Sect1.3.2} and example therein). This is a strong reason for
one to consider more than one field when studying distant galaxies. Generally an uneven
number of fields, from 3 to 7 is ideal to perform a redshift survey, allowing a
statistically valid check of cosmological variance effects.

\subsection{Depth of the photometry and selection effects related to surface-brightness}

Galaxies should be selected from a sufficiently deep photometric catalog. The photometric
catalog can be tested by the photometric completeness to the deepest counts in the
literature. It is generally accepted that the catalog should be complete down to
$\sim$1--2 \hbox{magnitudes} fainter than the limit used to perform the galaxy selection
in a redshift survey.

Doing otherwise can cause an important fraction of extended sources with relatively low
surface brightness to be missed, especially at high redshifts where cosmological dimming
effects become very important. Additionally, selecting more distant galaxies is
increasingly difficult as they are many more intrinsically faint foreground galaxies (see
Sec.~\ref{C1_Sect1.5}). These distant galaxies often need photometric and spectroscopic
measurements in \gls{NIR}, which are further affected by surface brightness dimming. The
light from extended distant galaxies is affected significantly by cosmological dimming
with a factor of $(1+z)^{-4}$ that applies to the surface brightness (bolometric)
luminosity of these objects. The strong redshift dependence of this effect can
significantly affect the fraction of light detected in the outer isophotes of a galaxy,
and can lead to galaxies being excluded from the sample. Cosmological dimming strongly
affects extended sources as well as the faint sources at large redshifts. Deep photometry
is therefore required to pre-select galaxies in an unbiased way before attempting to
determine their redshifts from spectroscopy (see Ex. {\it Photometric depth to recover
$z=0.9$ spiral galaxies}).

\begin{figure}[h!]
 \centerline{\includegraphics{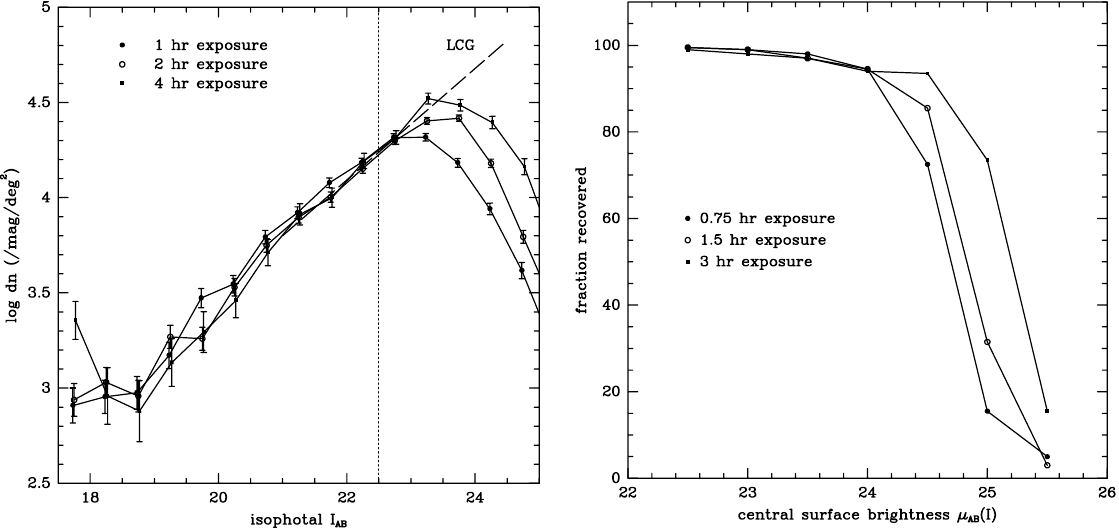}}
\caption[Photometry of distant galaxies and related surface brightness selection
effects]{Photometry of distant galaxies and related surface brightness selection effects
(from the \gls{CFRS} project, (Lilly et al., 1995), \copyright AAS, reproduced with
permission). {\it Left}: Number counts of objects with different exposure times. The
\gls{CFRS} reached down to $I_{AB}=22.5$ (assumed to be $I_{AB}$, vertical line) and all
source counts are complete to this limit, although only 4 h exposure (with \gls{CFHT})
allow a completeness to $I_{AB}=23.5$ (compare it to the dashed line that represents the
deepest counts). {\it Right}: Fraction of $I_{AB}=22.5$ galaxies that are recovered by
the \gls{CFRS} image detection algorithm, as a function of the central surface
brightness. A large exposure time (3--4~h) is necessary to recover $\ge$90\% of objects
with
$\mu(I)=24.5$ that correspond to a face-on, $L*$
exponential disk with a scale-length $h=6.6$kpc at $z=0.9$ and with a rest-frame
central surface brightness of $\mu_{0}(B)=21.5$.} \label{C1_Fig1}
\end{figure}

\begin{example}{Photometric depth to recover $z=0.9$ spiral galaxies}
Let us assume that most spiral-like galaxies at $z=0.9$ have a central surface
brightness similar to that of local spiral galaxies ($\mu_{0}(B)=21.5$).
Figure~\ref{C1_Fig1} shows that to recover most spiral-like galaxies up to $z\sim
0.9$ in a $I=22.5$ limited sample, it requires a photometry with a completeness up
to $I=23.5$. This also implies that detecting lower surface brightness galaxies
requires increasingly deeper photometry.
\end{example}

Distant galaxies are forming more stars than present day galaxies and their
intrinsic surface-brightness can be expected to be enhanced, making these objects
easier to detect. However, the fraction of low surface brightness galaxies in the
distant universe is unknown and these objects are most likely to be excluded from
a sample. It is therefore crucial to begin by requiring deep photometry as a
starting point if one wants to avoid severe biases while selecting a sample of
distant galaxies.

\subsection{Other selection effects$:$ Malmquist bias$,$ Eddington\\ bias$,$ and lensing
effects}

There are several gotchas that can negatively affect the quality of a sample of distant
galaxies. Even when attempting to select a sample of distant galaxies with a simple
criterion, there are several selection effects that can impact scientific results derived
from that sample. It is therefore necessary to carefully select the field, the limiting
magnitude and the sample size of a survey.


This also implies that one needs to avoid several selection criteria, as the
combination of the many related selection effects can be complex and can very
severely affect the representativeness and then the whole scientific outputs
derived from such a sample.

\begin{figure}[t]
 \centerline{\includegraphics[scale=0.95]{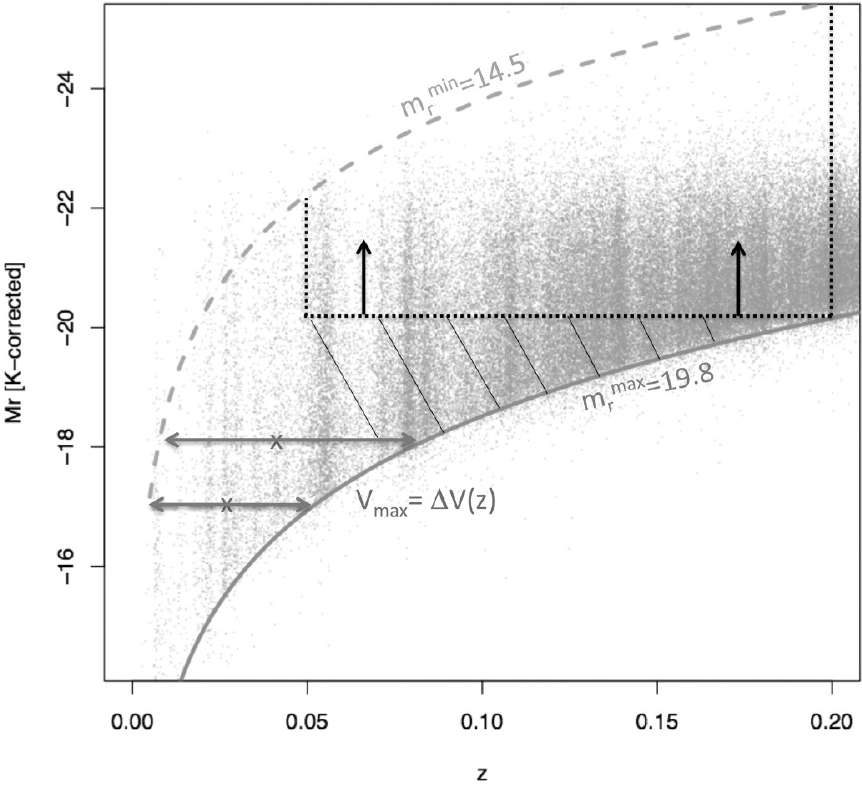}}
\caption[Selection of galaxies to avoid Malmquist bias]{\looseness-1Absolute,
$k$-corrected $M_r$ \gls{SDSS} magnitudes of galaxies (black points) selected from the
\gls{GAMA} survey as a function of redshift (adapted from (Rodrigues et al., 2015).
These absolute magnitudes are relatively good proxies for stellar mass. The solid and
dashed lines represent the \gls{GAMA} selection with $m_r\le 19.8$ and $\ge$ 14.5,
respectively. Selecting $0.05< z < 0.2$ galaxies between these two lines would cause an
important Malmquist bias, since for example, low luminosity galaxies with
$M_r\,{>}\,{-}20$ are not detected at $z=0.2$, while they represent most of the sample at
$z=0.05$. Avoiding such a bias requires selecting galaxies in the area delineated by the
dotted lines (see also the two vertical arrows), which eliminates the low luminosity
galaxies lying within the dashed area. The crosses represent two galaxies for which the
horizontal, double arrows indicate how $V_{\max}$ can be calculated for establishing the
LF (see Sec.~\ref{C1_Sect1.3.2} and Eq.~(\ref{C1_Vmax})).}\label{C1_Mr_z}
 \vspace*{-4pt}
\end{figure}

\vspace*{-7pt}
\subsubsection{Malmquist bias}

The Malmquist bias is probably the best known selection effect, although it
affects all average properties of galaxies, such as the luminosity density, and
therefore star formation or stellar density. A survey that is limited by a simple
apparent magnitude criterion would be unable to detect intrinsically faint
galaxies in its highest redshift bin, while those would be included in the sample
in the lower redshift bins. This is illustrated in Fig.~\ref{C1_Mr_z}, which also
indicates (see the dotted lines) how to prepare a sample not affected by the
Malmquist bias. Neglecting this may lead to significant errors in interpreting,
for example, scaling relations (see Chapter 5).

Naturally, calculations of the luminosity density evolution need to be corrected
for this effect. At a given redshift bin, one can use the immediately lower
redshift bin to evaluate the contribution of the galaxies that have absolute
luminosity just below the luminosity cut-off (see, e.g., Lilly et al., 1996). This
approach however requires a significant number of galaxies in relatively narrow
redshift bins to allow corrections that do not depend too much on evolution.

\setcounter{footnote}{3}

\subsubsection{Eddington bias$:$ magnitude errors near the magnitude limit}

Even the most simple and secure case of a magnitude limited sample can be affected by
subtle selection\vadjust{\pagebreak} effects. For example, consider a sample limited to
the magnitude of $m$, and for which the error in measuring the magnitude is $\delta m$.
It is often true that the number of fainter galaxies (those with $m+\delta m$) is larger
than the number of brighter galaxies with $m-\delta m$. For a random distribution of
errors between $\pm$$\delta m$, there will be more sources intrinsically fainter than the
limiting magnitude included in the sample than sources intrinsically brighter: this is
the Eddington bias. It may seriously affect a sample or a part of a sample where the
number count density is very steep.\footnote{This happens when $p>1$ for number count
density following $n\sim f_{\nu}^{-p}$.} This is the case for example with rare luminous
($L>\gls{L*}$) galaxies at the highest redshift bin of a survey, or with any bright
sources that are barely detected at large distances.

The Eddington bias can become important when combined with significant photometric
errors.

To recover the actual number count, Monte Carlo simulations accounting for the
distribution of the noise signal can be compared to the deepest observed counts
(see Ex. \emph{The Eddington bias for intermediate redshift LIRGs observed by ISO
in 1998}). This again shows the need for accurate magnitude measurements,
especially near the magnitude limit of the galaxy sample.

\begin{figure}[h]
 \vspace*{12pt}
 \centerline{\includegraphics{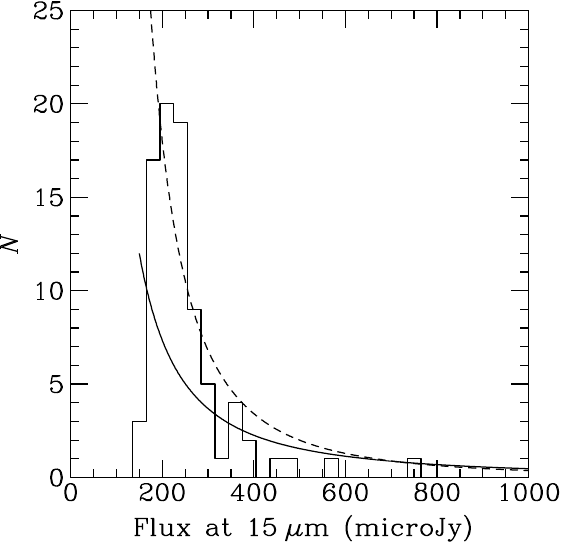}}
\caption[Number counts of ISO sources with $S/N > 3$]{The number counts of \gls{ISO}
sources with ${\rm S}/{\rm N}>3$ are shown by the histogram. The dashed line corresponds
to a slope $p=2.3$ $(n\sim f_{\nu}^{-p})$ fitting approximately the observed counts.
However, deeper observations showed that the slope should be $p=1.7$ (solid line). This
effect is due to the Eddington bias, and approximately 25\% of the sources in this
histogram are almost certainly below the ISO delection limit (250 microJy) and have
included into the sample as a result of their large flux calculations errors (from
Flores et al., 1999).} \label{C1_Fig2_Flores99}
\end{figure}

\begin{example}{The Eddington bias for intermediate redshift LIRGs
observed by ISO in 1998} The satellite \gls{ISO} detected for the first time the Luminous
InfraRed Galaxies (\gls{LIRG}s) at intermediate redshifts, \hbox{providing} the first
estimate of the density evolution of the star formation including both \gls{UV} and
\gls{IR} wavelengths (Flores et al., 1999). However \gls{ISO} was significantly less sensitive
than Spitzer and these \gls{LIRG}s were detected with \gls{S/N}$\ >4$ and $>3$ for the
complementary list. Although \gls{LIRG}s are much more abundant at intermediate $z$ than
locally, they are still rare sources and the slope of their number counts, as derived
using Spitzer, is $p=1.7$. However number counts from \gls{ISO} (see
Fig.~\ref{C1_Fig2_Flores99}) result in a steeper slope of $p=2.3$ because objects with
flux below 250 microJy are far more numerous than the brighter ones. This is therefore a
good demonstration of the effect of Eddington bias.
\end{example}

\subsubsection{Effect of gravitational lensing near the magnitude limit}

Gravitational lensing can also cause a somewhat similar selection bias when a galaxy
sample is obtained near the line-of-sight of a mass density such as a galaxy cluster or a
super-cluster.

Gravitational lensing affects the observed sources density since it causes a
brightening of background sources by $\delta \log(f_{\nu})= \log({\rm
magnification})$ (or $\delta m= -2.5\times \log({\rm magnification})$) while
reducing the area behind the lens due to the convergence of the light beam by a
factor equals to $({\rm magnification})^{-1}$. This results in a factor of
log($n/n_{0})=(p - 1)\times \log({\rm magnification})$, where $n$ is the observed
galaxy number density, $n_{0}$ is the source density if there were no lensing, and
$p$ is the actual slope of the number density counts. As it is the case with the
Eddington bias, gravitational lensing especially affects the brightest sources in
a sample (or sub-sample) since the latter often shows steep number counts.


\subsection{K-correction and choice of the selecting magnitude}\label{C1_Sect1.2.6}

Last but not least, the preparation of a redshift survey requires a careful choice of the
wavelength at which apparent magnitudes are used to perform the sample selection. As was
discussed in Sec.~\ref{C1_Sect1.1.1}, the choice of the selecting wavelength
$\lambda_{{\rm sel}}$ is crucial to (1) warrant a fair selection of galaxies
independently of their rest-frame color properties; (2) to accurately calculate the
rest-frame luminosities.

In fact, most redshift surveys have the goal of comparing the properties of
distant galaxies at different redshifts within the survey as well as with the
properties of nearby galaxies, where optical filters are defined from the $B$ to
the $R$ band and also more recently up to the $K$ band.

\begin{figure}
 \centerline{\includegraphics{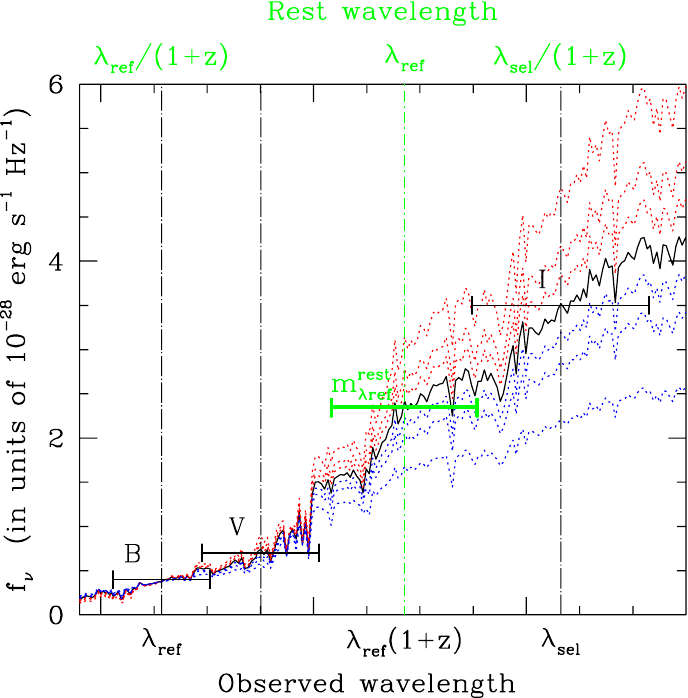}}
\caption[Method to calculate k-corrections]{Spectral energy distribution
$f_{\nu}$($\lambda$) of a \gls{CFRS} galaxy at $z= 0.5$, observed in $B$, V, $I$
broad-band filters (see horizontal solid bars). Fluxes ($f_{\nu}$) can be derived from
magnitudes using Eq.~(\ref{C1_AB}) and vice-versa. To derive $m^{{\rm
rest}}_{\lambda_{{\rm ref}}}$ (here the rest-frame $B$ magnitude at $\lambda_{{\rm ref}}=
457$nm, see the green, bold horizontal bar), one needs to adjust the broad-band
measurements using a spectrum (here a Sc type at $z=0.5$, see the solid line) extracted
from a grid of templates (here with types ranging from Sb to Sd, see red and blue dotted
lines). Rest-frame (observed) characteristic wavelengths are indicated on the top
(bottom) of the figure, respectively (see also the vertical dot-dash lines).
Equation~(\ref{C1_simplekcorr}) provides directly the $k$-corrected absolute magnitude
from $m^{{\rm rest}}_{\lambda_{{\rm ref}}}$.} \label{C1_kcorrect}
\end{figure}

Let us consider a redshift survey that is limited by a magnitude limit defined by
an observed filter centered at $\lambda_{{\rm sel}}$, i.e., including all galaxies
with $m_{\lambda_{{\rm sel}}}\le m^{\max}_{\lambda_{{\rm sel}}}$. For each galaxy
in the survey, a common rest-frame absolute magnitude at a reference wavelength,
$\lambda_{{\rm ref}}$, can be computed as:
\begin{equation}\label{C1_absmag}
M_{\lambda_{{\rm ref}}}=m_{\lambda_{{\rm sel}}}- 5\times \log(d_{L}(z))-k,
\end{equation}

\noindent where $d_{L}(z)$ is the cosmological luminosity distance and $k$ is the
$k$-correction. When a source is observed at a redshift $z$ in a filter centered at
$\lambda_{{\rm sel}}$, this provides a direct estimate of the rest-frame absolute
magnitude at $\lambda_{{\rm sel}}/(1+z)$, a wavelength that is generally different than
the desired $\lambda_{{\rm ref}}$ (see Fig.~\ref{C1_kcorrect}).

The $k$-correction $k$ has to be computed to account for this discrepancy. If a
galaxy happens to be at a redshift that is exactly $z=\lambda_{{\rm
sel}}/\lambda_{{\rm ref}}-1$, one can compute $M_{\lambda_{{\rm ref}}}$ and the
$k$-correction is simply $k= -2.5\times \log(1+z)$ to account for the fact that
the luminosity $L_{\nu}$ (or the flux $f_{\nu}$ that is related to AB magnitude,
see Eq.~(\ref{C1_AB})) is not bolometric but a density per unit
frequency\footnote{This assuming the energy conservation ($E=\nu \times
f_{\nu}=\nu_0 \times f_{\nu_{0}}$), which leads to a flux at rest $f_{\nu_{0}}$
that is equal to $f_{\nu}/(1+z)$.} (Hogg et al., 2002)

In most cases, however, {\it a priori} knowledge of the galaxy spectral energy
distribution (\gls{SED}) is required to compute the $k$-correction. The exact, and
complex definition of the $k$-correction can be found in (Hogg et al., 2002).

In practice, an estimate of the $k$-correction can be obtained with a limited
knowledge of the \gls{SED} that only samples the rest-frame wavelength range from
$\lambda_{{\rm ref}}$ to $\lambda_{{\rm sel}}/(1+z)$. Estimating $k$-corrections
is simpler if all of the galaxies have actually been observed at $\lambda_{{\rm ref}}$
in addition to $\lambda_{{\rm sel}}$, providing the very useful ($m_{\lambda_{{\rm
ref}}}-m_{\lambda_{{\rm sel}}}$) observed color (see Fig.~\ref{C1_kcorrect}). The
$k$-correction is then:
\begin{equation}\label{C1_kcorrection}
k= -2.5 \times \log(1+z) - (m^{{\rm rest}}_{\lambda_{{\rm ref}}} -
m_{\lambda_{{\rm sel}}})
\end{equation}
and Eq.~(\ref{C1_absmag}) simplifies to:
\begin{equation}\label{C1_simplekcorr}
M_{\lambda_{{\rm ref}}}=m^{{\rm rest}}_{\lambda_{{\rm ref}}} - 5\times
\log(d_{L}(z)) + 2.5\times \log(1+z),
\end{equation}

\noindent where $m^{{\rm rest}}_{\lambda_{{\rm ref}}}$ is the rest-frame magnitude that
the galaxy would have at $\lambda_{{\rm ref}}$ in the rest frame.
Figure~\ref{C1_kcorrect} shows how it can be interpolated from the ($m_{\lambda_{{\rm
ref}}}-m_{\lambda_{{\rm sel}}}$) color, which is used to assign an \gls{SED} type to each
galaxy, using a grid of colors obtained from redshifted galaxy synthesis models.

Other colors can be used to interpolate $m^{{\rm rest}}_{\lambda_{{\rm ref}}}$ and
the $k$-correction, but they should be defined over a wavelength range that
includes\footnote{For example, the $V-I$ color may be used to interpolate the
(reference) $B$ mag from $z=0.2$ to 0.8 (see Fig.~\ref{C1_kcorrect}).}
$\lambda_{{\rm ref}}\times(1+z)$. It is however strongly recommended to avoid any
extrapolation that can lead to considerable errors in estimating
\hbox{$k$-corrections}.


For samples of galaxies spanning a small redshift range, the associated errors can
be very small ($\sim$0.05 mag) by setting $\lambda_{{\rm ref}}$ accordingly to the
median redshift ($z_{{\rm med}}$), i.e., $\lambda_{{\rm ref}} \sim \lambda_{{\rm
sel}}/(1+z_{{\rm med}})$~(Hammer et al., 2001). The calculations of magnitudes, colors
and the use of \gls{SED} libraries are described in Chapter 2.

Old stellar populations (F, G, and K) have a strong signature at 400nm (the 400nm
break) causing a strong decrease in luminosity at $\lambda < 400$nm (see
Chapter~2). This signature has to be accounted for when defining $\lambda_{{\rm
sel}}$ in redshift surveys that aim at comparing distant galaxy properties with
the properties of local galaxies. This allows for a better determination of the
stellar mass of a galaxy as the galaxy mass can be dominated by old stars emitting
most of their light above 400nm rather than by young and blue stars with low
$M_{{\rm stellar}}/L$ ratio (see Ex. \emph{Selecting wavelength for comparing
$z=0$ to $z=0.8$ galaxies}).

\begin{figure}[h!]
 \centerline{\includegraphics{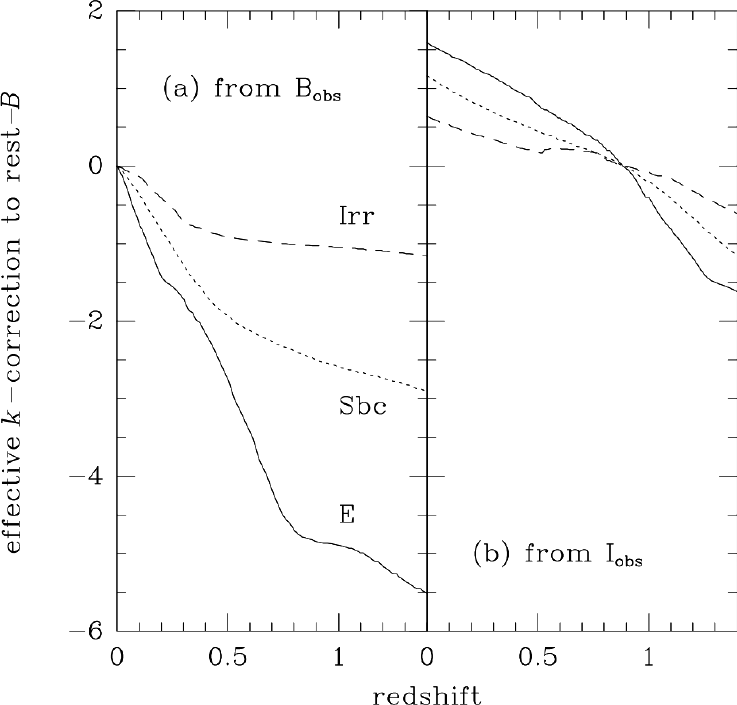}}
\caption[k-corrections for 3 galaxy types]{Variations of the $k$-corrections in magnitude
(from Lilly et al,., 1995), \copyright AAS, reproduced with permission) for three types of
galaxies, E, Sbc, and Irr, assuming the \gls{LF} is computed in absolute magnitude in $B$
band ($\lambda_{{\rm ref}}=457$nm). {\it Left}: Assuming that galaxies have been selected
from their observed B band magnitudes ($\lambda_{{\rm sel}}=457$nm). {\it Right}:
Assuming that galaxies have been selected from their observed I band magnitudes
($\lambda_{{\rm sel}}=832$nm).} \label{C1_Fig2}
\end{figure}

\begin{example}{Selecting wavelength for comparing $\bm{z=0}$ to $\bm{z=0.8}$ galaxies}
Figure~\ref{C1_Fig2} compares the $k$-corrections of a survey aiming at
characterizing the rest-frame $B$ band properties of galaxies from $z=0$ to 0.8.
If galaxies are selected in the observed $B$ band, with increasing redshift this
band progressively samples the rest-frame ultraviolet light (see, e.g.,
Fig.~\ref{C1_kcorrect}). This leads a very wide range of $k$-corrections within
the different \gls{SED} types, that are up to 4 mag at $z=0.75$. This results in
large uncertainties, especially since the properties of the different galaxy types
are not very understood in the \gls{UV}.

In contrast, a selection in the $I$ band is an exact match to the rest-frame $B$
band at $z=0.82(\lambda_{{\rm ref}}=\lambda_B = \lambda_{I}/(1+z) = \lambda_{{\rm
sel}}/(1+z))$. The ($B-I$) color (equivalent to the ($m_{\lambda_{{\rm
ref}}}-m_{\lambda_{{\rm sel}}}$) color) allows us to interpolate with a far better
accuracy and results in much smaller $k$-corrections all the way from $z=0$ to
$z=0.82$, because $\lambda_B \le \lambda_I/(1+z) \le \lambda_I$.
Figure~\ref{C1_Fig2} illustrates that by selecting distant galaxies using the
apparent $B$ band magnitude, one would exclude elliptical galaxies at high
redshift, since these galaxies are made of old and red stellar populations with a
low continuum emission in the \gls{UV}. The $I$ band selection allows for a much
more representative sample of galaxies up to $z=0.82$.
\end{example}


This further demonstrates how a selection in the $I$ band ($\lambda_{{\rm sel}}=832$nm)
is the best way to characterize the rest-frame $B$-band properties for a survey of $z\le
0.8$ galaxies.

Surveys at $z\le 2, 3$, and 4 require a selection done in the $J$ ($\lambda_{{\rm
sel}}=1200$nm), $H$ ($\lambda_{{\rm sel}}=1600$nm) and $K$ ($\lambda_{{\rm
sel}}=2100$nm) band, respectively.

When possible, it is even better to select objects in even redder filters to
sample the rest-frame properties of galaxies at wavelengths that are not strongly
affected by star formation and dust extinction, such as the rest-frame $R, I$, or
$K$ bands.

\chapter{Imaging and Photometry}
{\it Here it is an excerpt of Chapter 2.}

\section{\label{C2_morpho}Galaxy Morphology}

\subsection{\label{C2_morphotree}A pragmatic and conservative approach
  to classify distant galaxies}

Classifying the morphology of distant galaxies is usually with the goal to study
galaxy evolution. In order to do this, one must assume that distant galaxies are
indeed similar to nearby galaxies, i.e., that their morphologies can be
characterized using a similar sequence as in the Local Universe. This conservative
assumption is the only one allowing to properly calibrate morphological evolution,
and has been demonstrated to be robust through comparison with kinematic studies
(e.g., Neichel et al., 2008, Hammer et al.,09).

\setcounter{box}{7}

\begin{example}{\label{C2_ClassifTree}How to classify a galaxy morphology using a semi-empirical decision tree{\em \textbf{?}}}
\begin{enumerate}
\item Extract stamp and color images (see Ex. \emph{How to construct a
    color image}?) for all galaxies to be classified;
\item Derive $R_{{\rm half}}$ for all galaxies to be classified (see Ex.
  \emph{Measuring the surface brightness profile and basic properties
    of a distant galaxy});
\item Determine the compactness criterion: there is a limit below which
  galaxies are too small to be reliably analyzed because of the limited
  spatial resolution of the images. A strict \hbox{lowest}
  limit to the half-light radius is given by the \gls{PSF} size, i.e.,
  $R_{{\rm half}} \sim 0.5\times {\rm FWHM}$.
  However, because the measurement process of the other steps introduce an
  additional intrinsic scatter, the real limit is larger.  This limit must therefore
  be derived empirically, by testing the whole decision tree on a test
  sample. In the case of $z\sim 0.6$ galaxies observed with \gls{HST}/\gls{ACS}
  images, (Delgado-Serrano et al., 2010) found that this limit is $R_{{\rm half}}
  \sim 1.33\times {\rm FWHM}=1$ kpc, corresponding to 0.15 arcsec;
\item Isolate compact galaxies in the sample using the compactness
  criterion;
\item Decompose all of the remaining galaxies into disk and bulge
  (e.g., using \gls{GALFIT}, see Ex. \emph{How to fit a galaxy with
    \gls{GALFIT}}?);
\item Assess the bulge/disk decomposition reliability using the
  residual map and profile cut. Examine whether residuals show
  asymmetric and/or multiple components (see Fig.~\ref{C2_FigClassifTree}). Check for off-centered bulges with respect
  to the disk, i.e., whether the bulge and disk component are
  consistent;

\item With each reliable fit, derive the $B/T$ flux
  ratio (e.g., as given by \gls{GALFIT}) and isolate sub-classes depending
  on the $B/T$ value (see Fig. \ref{C2_FigClassifTree});
\item With each sub-class, examine the color maps to check  whether all
  peculiar sub-structures identified in the residual map and/or stamp
  image are associated with physically distinct components. In cases of
  reliable bulge/disk decomposition, check whether a central bulge (if
  present) is redder than the disk, which is necessary to be able to classify a
  galaxy as a spiral. This is similar to the appearance of spiral galaxies in
  the local Universe and ensures that galaxies classified as spirals
  are indeed comparable in terms of color gradient at any redshift;
\item Based on the result of the last step, assign a final morphological
  class to the galaxy;
\item Classify the whole sample using at least three different
  classifiers and compare the results;
\item Repeat the classification until all classifiers agree on the
  final morphological class, while keeping track of the initial rate of
  disagreement, which will give an order of estimate of the
  classification accuracy.
\end{enumerate}
\end{example}


\setcounter{figure}{9}

\begin{figure}[!h]
 \centerline{\includegraphics{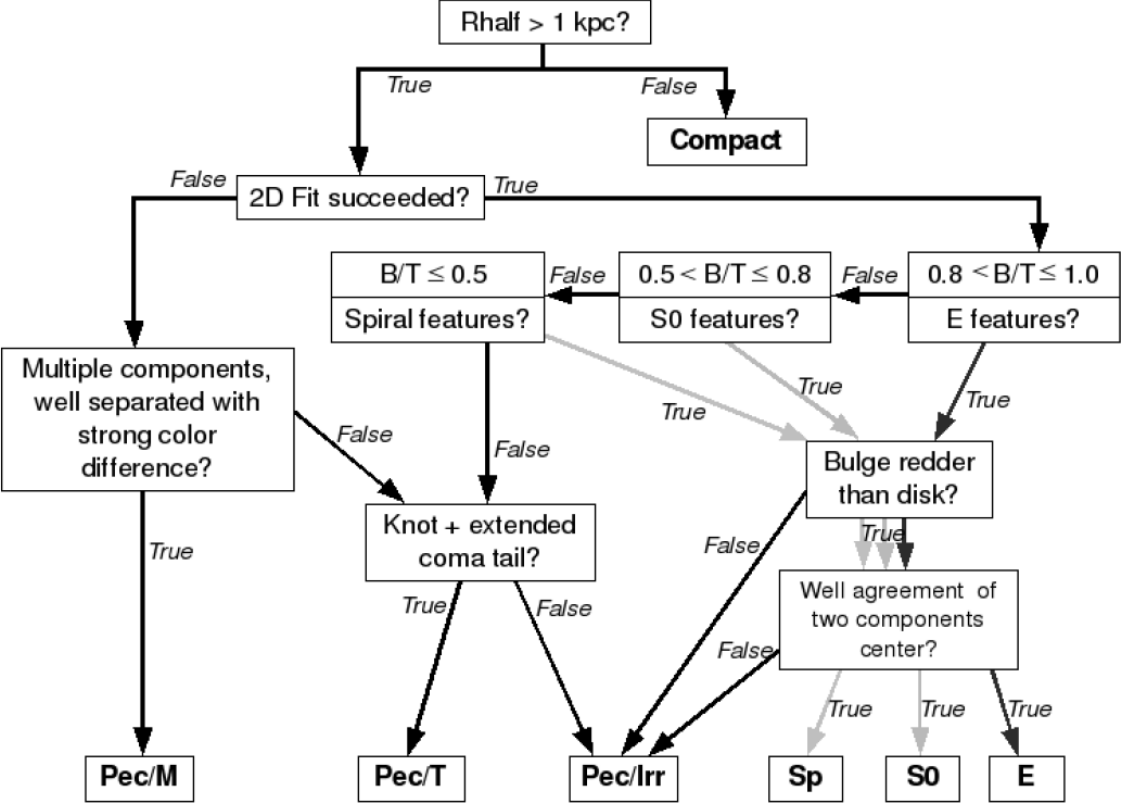}}
\caption[Morphological classification tree]{Semi-Empirical
  morphological classification tree defined by
  Delgado-Serrano \emph{et~al.} (Delgado-Serrano et al., 2010) to classify the morphology of $z\sim 0.6$
  galaxies. Credit: Delgado-Serrano et al., 2010, A\&A, 509, 78,
  reproduced with permission \copyright ESO.\label{C2_FigClassifTree}}
\end{figure}

Such a method was first developed by (Zheng et al., 2004, Zheng et al., 2005, Neichel et al., 2008) for spiral
galaxies, and extended by (Delgado-Serrano et al., 2010) to early-type galaxies. It is a
semi-empirical method that relies on visual inspection of images using a well defined
sequence of procedures in order to  limit the subjectivity and improves the
reproducibility of the classification. The sequence of procedures is clearly defined and
ordered according to a decision tree that must be followed every time a galaxy morphology
is being classified (see Ex. \emph{How to classify a galaxy morphology using a
semi-empirical
  decision tree}?). Specific decision trees must be created based on what observables
are available for the classification. As the $B/T$ light ratio is one of the most
correlated parameter to the Hubble sequence (e.g., De Lapparent et al., 2011), this
parameter should  be at the core of any semi-empirical classification, as illustrated
in\break Fig.~\ref{C2_FigClassifTree}.

A decision tree based approach requires the following preliminary steps:
\begin{enumerate}
\item Ensure that all images are deep enough to avoid surface
  brightness effects and sufficiently resolved to limit the fraction
  of compact galaxies that are not properly resolved;

\item Ensure that all quantities ($B/T$ and $R_{{\rm half}}$) are defined in
  an optical rest-frame (i.e., at wavelengths larger than $\sim$500nm) as well as the decomposition of the galaxy light profile to estimate
  residuals, and scale lengths of the  bulge and of the disk components;
\item Prepare at least one color map (see Ex. \emph{How to construct a
    color image of a galaxy}?).
\end{enumerate}

It is only by combining all these different pieces of information and by comparing
the classifications established by three experienced astronomers that
one can achieve a robust galaxy morphological classification. Even
simpler classifications compared to Fig. \ref{C2_FigClassifTree} can
be obtained by merging all peculiar galaxies in a single category,
i.e., Peculiar. Indeed, it can be relatively difficult to detect
merging features such as tidal tails because they rapidly vanish
within less than few hundred years after the merger, they depend on
the exact merger orbit, and they correspond to extended and
low-surface brightness features strongly affected by cosmological
dimming.

The decision tree (see Fig. \ref{C2_FigClassifTree}), especially in the simplified
form described above, has been shown to lead to a morphological classification
that is in good agreement with (1) manual classification made by world experts in
galaxy morphology (Van den Bergh 2002) and (2) with a fully independent kinematic
classification (Neichel et al., 2008). While the resulting classification is limited by
the relatively small size of the sample, it leads to a new and more modern galaxy
classification scheme, i.e., a morpho-kinematic classification (Hammer et al., 2009
and see Sec.~\ref{C4_morphokin}).

\subsubsection{\label{C2_Sectcolormaps}Color maps}

Magnitude-scale color-images can be created by  combining two images in different
bands (Abraham et al., 1999, Menanteau et al., 2001, Zheng et al., 2004). A color image is  only useful when
the spatial resolution of the original images
 is similar. Color maps are clearly limited to objects that
are properly resolved, i.e., with sizes larger than 2--3 times the \gls{PSF} size.
For distant galaxies, such images should be constructed using space telescope
images. Since the integration times of
 the respective band images are not necessarily the same, a \gls{S/N} image
must also be generated to identify  reliable pixels in the color space (see
Zheng et al., 2004 and Ex. \emph{How to construct a color map of a
  galaxy}?). The advantage of the Zheng et al. (2004) method is that it
allows one to also estimate colors in fully extinguished regions. Color maps have a wide
range of applications, from pixel-by-pixel \gls{SED} analysis (e.g.,
Abraham et al., 1999, Menanteau et al., 2001, Zibetti et al., 2011), to morphological classification
(Zheng et al., 2004; Delgado-Serrano et al., 2010;\break see Fig.~\ref{C2_ColorMap}).

\setcounter{footnote}{22}

\setcounter{box}{8}

\begin{example}{How to construct a color map{\em \textbf{?}}}
\label{C2_ExColorImage}
\begin{enumerate}
\item Construct two stamp images with the same physical size in the two
  filters $X$ and $Y$. You will need images that
  include the sky
  background. Estimate their spatial resolutions FWHM (see Ex.:
  \emph{How to estimate the spatial resolution of an image}?). If for
  instance the spatial resolution in $X$  is larger than in $Y$, then
  convolve the  $Y$ image using a Gaussian kernel of size
  $W=\sqrt{{\rm FWHM}_{X}^2-{\rm FWHM}_{Y}^2}$. The \gls{IRAF} task
  \emph{PSFmatch} can be a useful tool for determining the
  kernel W;
\item Determine the spatial shift and rotation between the two images
  using at least three stars located as close as possible to the
  target galaxy and align the two images. Both images should be aligned
  down to a precision that is smaller than one pixel in both directions;
\item Construct the color map $\mu_{X-Y}$ defined as\footnotemark:
  $$\mu_{X-Y}=-\frac{1}{2}\log{\left(\frac{\sigma^2_X}{F^2_X}+1\right)}+
  \frac{1}{2}\log{\left(\frac{\sigma^2_Y}{F^2_Y}+1\right)}+\log{(F_X/F_Y)}$$
  The associated noise within each pixel can be calculated using Eq.
  (10) of Zheng et al. (2004):
  $$\sigma_{X-Y}^2=\log{\left(\frac{\sigma^2_X}{F^2_X}+1\right)}+
  \log{\left(\frac{\sigma^2_Y}{F^2_Y}+1\right)}$$
  Here, both the signal and noise include \emph{both} the source and
  background. These relation are based on the property that
  $\sigma_X^2$ and $\sigma_Y^2$ are dominated by Poisson noise, which are close
  to being Log-normal distributions, which therefore allows one to estimate
  the associated \gls{S/N} map as $\mu_{X-Y}/\sigma_{X-Y}$.
\item Threshold the color map using a relative threshold in \gls{S/N}
  to select reliable pixels, e.g., those with a \gls{S/N} larger than
  four times the mean \gls{S/N} in the background.
  \end{enumerate}
\end{example}

\footnotetext{Note that the corresponding Eq. (9) of Zheng et al., (2004)
  misses a square factor in the denominators, and that $\mu_{X-Y}$ is
  positive for Poisson statistics with at least two counts.}

\setcounter{figure}{10}

\begin{figure}[!h]
 \centerline{\includegraphics{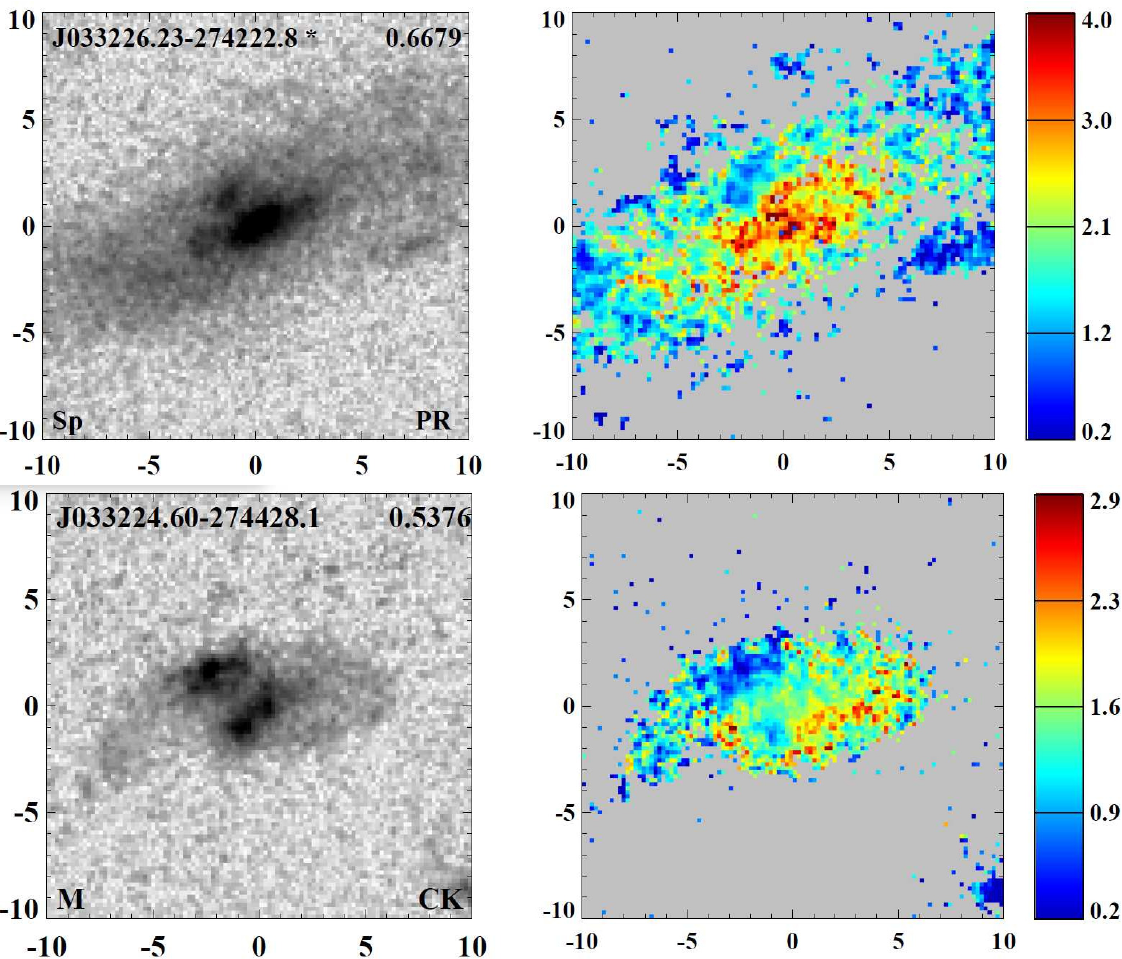}}
\caption[Examples of color maps of z$\sim$0.6 galaxies.]{\looseness1Examples of
  $B$--$z$ color maps for two $z\sim0.6$ galaxies from the \gls{IMAGES}
  survey. Credit: Neichel et al. 2008, A\&A, 484, 159, reproduced with
  permission \copyright ESO. \emph{Top panel:} J033226.23-274222.8,
  which was classified as a spiral galaxy. The inner red ellipse
  corresponds to $\sim R_{{\rm half}}$ with $B-z=2.2$, while the outer blue
  ellipse corresponds to $\sim 2.R_{{\rm half}}$ with $B-z=1.8$. The $B-z$
  color between $R_{{\rm half}}$ and $2.R_{{\rm half}}$ is found to be $\sim$1.5.
  This difference of color as a function of aperture size, typical of
  $z\sim 0.6$ spirals (see [Fig. 10] Neichel et al., 2008), can
  significantly bias the photometric aperture corrections (see Ex.
  \emph{Biases in colors caused by uncontrolled aperture
    corrections}); \emph{Bottom panel}: J033224.60-274428.1, which was
  classified as a peculiar galaxy because of the strong color gradient
  between the two brightest components, which could be indicative of
  the two progenitors of an on-going merger.\label{C2_ColorMap}}
 \vspace*{60pt}
\end{figure}

\chapter{Integrated Spectroscopy}
{\it Here it is an excerpt of Chapter 3.}
\section{Emission lines}\label{C3_Emission_lines}

\subsection{Proper methods for measuring emission lines}
\label{C3_Emission_lines_measures} Before measuring the width or flux of the lines, the
observed spectrum has to be set to the rest-frame. This can be done by dividing the
wavelength scale by $(1+z)$ and multiplying the fluxes $F_\lambda$ by $(1+z)$. Notice
that $F_\nu$ should be divided by $(1+z)$.

At low and moderate resolution, emission lines often display profiles compatible
with a Gaussian distribution. The line is measured by fitting a simple Gaussian
model. In galaxy spectra, lines lie above the continuum produced by the stars, and
a continuum component is added to the Gaussian model, usually a low polynomial
order (a~constant or a line). The flux of the line is measured by integrating the
spectra between $\pm3\sigma$ of the central wavelength of the fitted Gaussian (see
Fig. \ref{C3_Fig_Emission_iraf}).


For helium and hydrogen recombination lines, such as the Balmer lines, the underlying
continuum is more complex because the emission lines are superimposed on the absorption
lines produced by the same transition in stellar photospheres (see Fig.
\ref{C3_Fig_Spectra_with_continum} and Example \textit{Correct from underlying
absorption}). Not accounting for the underlying absorption when measuring the emission
lines leads to the \hbox{important} underestimation of their fluxes and thus induces a
bias for all the derived properties (Liang et al., 2004a).

\vfill\eject

\setcounter{figure}{13}

\begin{figure}[h!]
 \centerline{\includegraphics{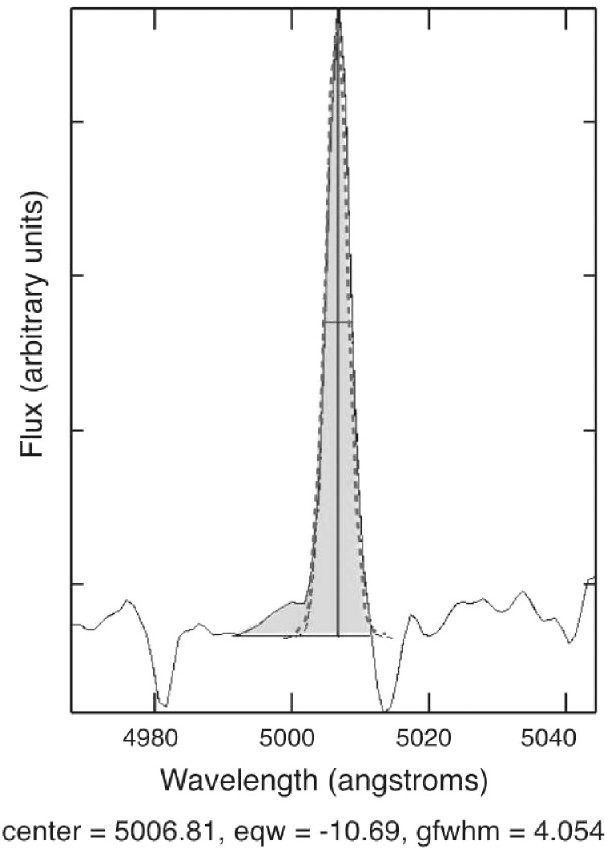}}
\caption[Measure emission lines with {\sc iraf}]{Measure of the
  [\ion{O}{3}]5007 emission line in a $z=0.6$ galaxy spectrum, using
  \gls{IRAF}. The line is composed of a main component at the
  systematic velocity of the galaxy (dashed line), and a blue shifted
  broad component. The total emission flux from both components can be
  measured with \gls{IRAF} by integrating the spectra in the vicinity of
  the two components (filled area). }
\label{C3_Fig_Emission_iraf}
\end{figure}

\setcounter{box}{4}

\begin{example}{Correcting from underlying\\ absorption}
\textit{The most efficient method to subtract the stellar component consists in fitting a
synthetic spectrum from a template library to the spectrum of each galaxy. The {\sc
starlight} software is taken for illustration but the steps are valid for all full
spectrum fitting softwares. For correcting the underlying absorption in Balmer emission
lines the steps are}:

\medskip (1) \emph{Adjust the spectral resolution between the templates and the observed
spectrum.} A majority of softwares automatically match the spectral resolution between
the spectral templates and the input spectrum. This works reasonably well in high
\gls{S/N} spectra and when the difference between the~two spectral resolution is not too
large (Cid Ferandes et al., 2005). However, for \gls{S/N} in the continuum under ten, it is
highly recommended to adjust the template spectral resolution prior to that of the
observed spectra, using a Gaussian deconvolution. Always use templates with higher
spectral resolution than the observations. Mismatch on the resolution induces bias in the
fluxes of Balmer lines.

\medskip (2)
\emph{Mask emission lines in the spectrum to be fitted.} Regions associated with emission
lines have to be excluded from the fit. In {\sc starlight} this can be done by setting
the weight flag to 0 in the input file. Special attention has to be taken when excluding
the emission in Balmer lines: the wings of the absorption line, not contaminated by the
emission line, should not be masked since these features give strong constraints on the
fit. Strong sky emission or absorption also need to be excluded, see the grey vertical
lines in Fig. \ref{C3_Fig_Spectra_with_continum}.

\medskip (3)
\emph{Create a library base.} The measurement of the fluxes does not strongly depend on
the choice of template library. In general, the stellar evolutionary population models
are used as base templates in the literature because of their completeness in the $Z$-age
grid and their wide wavelength coverage, but empirical stellar or star cluster libraries
can also be used. The only requisite is that the templates spectra cover the wavelength
range of the input spectrum and that the spectral resolution is higher than the observed
spectrum.

\medskip (4) \emph{Fit and quality control.} The fitting algorithm will find the best mix of
stellar populations from the template base that adjusts the observed
spectrum.\footnotemark The quality of the fit is evaluated by the $\chi^2$ value and a
visual inspection of critical features, such as the wings of the Balmer lines, the
4000\AA, break and the G band, as well as the global slope of the synthetic spectrum. If
the slopes of the observed and synthetic spectra do not match, check the
spectrophotometric quality of the input spectrum or wherever a nebular continuum should
be added in the template base for extremely young objects.

\medskip (5) \emph{Subtract the synthetic spectrum to the observed spectrum}. The synthetic
spectrum is subtracted directly to the observed spectrum and emission lines can be
measured in the continuum-subtracted spectrum. The continuum of the final spectrum should
be flat and close to 0.\vadjust{\pagebreak} In particular, check the quality of the fit
close to emission lines. A negative continuum means that the absorption lines have been
underestimated, while a bump suggests that the underlying absorption has been
overestimated. Should one of these problems arise, check that: (a) the spectral
resolution has been correctly matched between the observed spectrum and the templates;
(b) the Balmer emission lines are correctly masked; (c) if the problem remains this can
be due to the template base used, e.g., a young stellar template may be missing.

\medskip

The estimation of the synthetic spectrum is sensitive to the wavelength window available
for the fit. For example, the spectral window between [\ion{O}{2}]$\lambda$3727 and
[\ion{O}{3}]$\lambda$5007 has several strong absorption lines (e.g., Balmer lines and Ca
lines) and the 4000\AA, break, which give strong constraints on the fit. The underlying
absorption in Balmer lines can thus be accurately constrained when fitting this
wavelength interval.
\end{example}

\setcounter{footnote}{8}

 \footnotetext{At this point, we strongly advise against using the fit
  results to formulate any conclusion about star formation history,
  extinction, or stellar mass. The reader is referred to Sec.~\ref{C3_PropertiesStellarPopulations} for a
  complete discussion on stellar population decomposition.}

\begin{figure}[h!]
 \centerline{\includegraphics{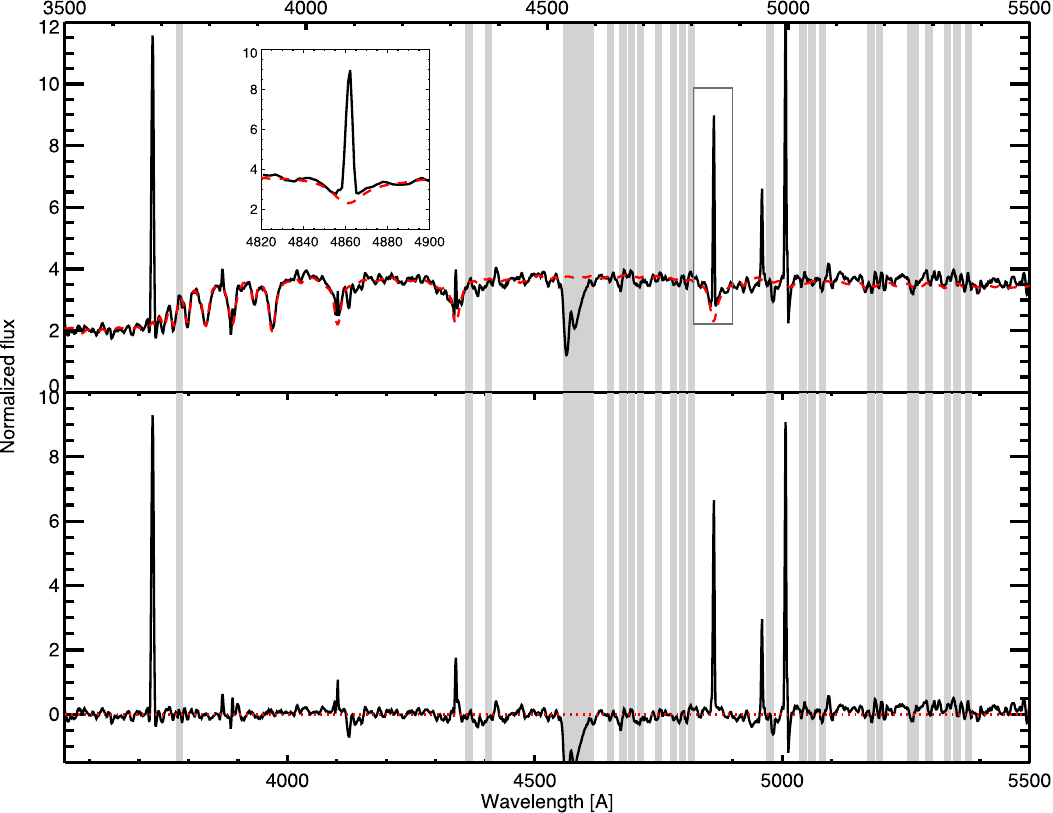}}
\caption[Correcting from underlying absorption]{ Correcting from underlying absorption in
a spectrum of a $z=0.66$ galaxy. \emph{Up}: The continuum of the rest-frame spectrum
(black line) has been fitted using a stellar library and the {\sc starlight} software.
The contribution from the stellar absorption is plotted in red line. The insert box is a
zoom on the H$\beta$ line. The emission line and the underlying absorption are clearly
visible in the H$\beta$ line. The regions affected by sky lines and the strong O$_2$
absorption feature from the atmosphere (at 7600\AA, in the observed frames) are indicated
by grey areas and have been masked during the fit. \emph{Down}: Final spectrum following
subtraction of the stellar continuum, from which the flux of the emission lines can be
properly measured. } \label{C3_Fig_Spectra_with_continum}
\end{figure}

There are many softwares designed to automatically measure emission lines in large
number of spectra and for which the stellar continuum subtraction is included.
They are generally based on the assumption that emission lines have simple
Gaussian profiles. Since this is not always the case and that at increasing
redshifts, galaxies have more complex and disturbed kinematics
(Yang et al., 2008), such measurements can lead to systematics
($\sim$10\%). Moreover the Poisson noise can affect the profile of faint emission
lines, which may become difficult to detect with automatic algorithms. It is thus
recommended to  manually measure the lines, or at least to manually check the fit
accuracy during this process.

\pagebreak

\subsection{Low S/N regime$:$ measurement bias}\label{C3_Emission_lowSN}

In most distant galaxy spectra, some of the diagnostic lines are faint and with
low \gls{S/N}. Here, we show why this problem has to be treated and provide a
solution for emission line ratios as well as limits on undetected emissions. The
\textit{profile signal-to-noise}, can be defined as the median \gls{S/N} in a
line:
\begin{equation}
{\rm S}/{\rm N}_{{\rm line}} = \frac{F_{{\rm line}}}{ \sigma _c \times \sqrt{N}},
\end{equation}
where $F_{line}$ is the line flux, $\sigma _c$ is the noise of the
pseudo-continuum, $N$ the number of resolution elements over which the line
spreads. If the line has a Gaussian profile, $N$ is equal to six times the
emission line width.

The statistical significance is generally evaluated as a function of the equivalent
fluctuation of a standard Gaussian stochastic variable. A \gls{S/N} above 3 implies that
the line is detected within a $3\sigma$ significance level, and thus with a probability
above 96\%. However, the $3\,\sigma$ significance level criterion assumes that the
detected signal lies above a known background, affected with pure Gaussian noise. These
assumptions are not valid in the case of galaxy spectra, since the continuum is not known
{\it a priori}, and it is not constant as a function of wavelength due to the underlying
stellar populations. Even after subtracting the stellar continuum (see Ex.
\textit{Correct from underlying absorption}), the background still has residual signals
arising from uncertainties in the flux calibration and templates\break mismatches.

In an ideal case of Poisson noise and a Gaussian line shape with unknown
background, Rola et al., (1994) demonstrated that the integrated flux and \gls{S/N}
of a narrow emission line can be described by a lognormal probability
distribution. This has a strong impact in the low S/N regime:
\begin{itemlist}
\item The fluxes from lines having observed \gls{S/N} under 5 are systematically
overestimated,  because \textit{``At very low \gls{S/N} an emission line can be
detected with the help of some chance positive flux fluctuation, but never if it
is a negative fluctuation. Most of the time, the observed line intensity is the
true line intensity plus a noise intensity under the line, thus the bias''};
\item  Observed \gls{S/N} over 4--5 (respectively  $2\sigma$ and $ 3\sigma$ confidence level) are required to exclude false detections;
\item As for color in imagery (see Chapter 2, Sec. \ref{C2_Sectcolormaps}), the
resulting noise of the logarithm of a line ratio follows a normal distribution
(see Appendix in Zheng et al., 2004).
\end{itemlist}

\begin{figure}[!b]
 \centerline{\includegraphics{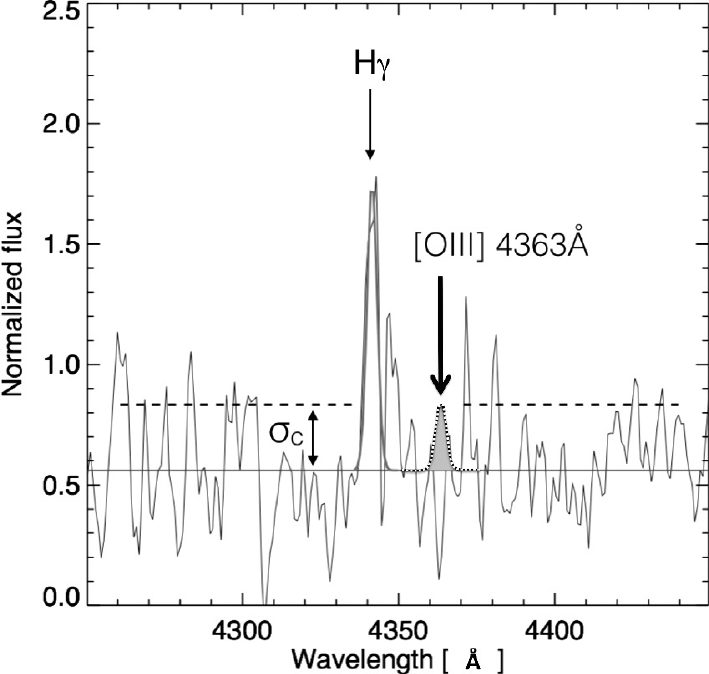}}
\caption[Flux limit of a non-detected line]{{Portion of a \gls{FORS2} spectrum of a
$z\sim 0.6$ galaxy, which is shown at rest-frame. The goal is to estimate a limit on the
[O \textsc{iii}]$\lambda4363$\AA\ line while $H\gamma$ is detected and can be used to
estimate the expected width of a line (${<}\delta _v{>}$). One has to evaluate the noise
in the pseudo-continuum ($\sigma _c$), and from the Gaussian surface (grey area) one may
derive the maximal line flux possessed by a line at 4363\AA\ above the continuum (see the
Gaussian with a dotted line) represented by the horizontal full line.}}
\label{C3_Flux_limit}
 \vspace*{-6pt}
\end{figure}

Estimating the \emph{upper limit} of the flux of an undetected line can be very useful,
including to set limits on line flux ratios. Figure~\ref{C3_Flux_limit} shows how this
can be done by assuming that the undetected line has a Gaussian line-profile with a width
equals to the mean width (${<}\delta _v{>}$) of other detected lines, and a height equals
to the pseudo-continuum noise $\sigma _c$. The line-integrated flux is then $F
=\sqrt{2\pi}\cdot\sigma _c \cdot\,{<} \delta _v{>}$, where $\sqrt{2\pi}$ is coming from
the integration of the Gaussian profile, $\sigma _c$ is in flux units (either $f_{\nu}$
or $f_{\lambda}$), and ${<}\delta _v{>}$ is either in Hertz or Angstrom, respectively.

\chapter{Integral Field Spectroscopy}
{\it Here it is an excerpt of Chapter 4.}

\section{Data reduction of  IFU observations}\label{C4_datared_3D}

\subsection{Optimizing sky subtraction with NIR IFUs}

In the \gls{NIR} range, residuals from the subtracted skylines are so intense that
they can still affect the detection of target emission lines (see
Sec.~\ref{C3_Prepare_Sky}). The target has to be carefully chosen in redshift so
as to avoid contamination of the galaxy emission lines by strong skylines,
otherwise the emission line moments cannot be measured. Figure~\ref{C4_figSkyKMOS}
demonstrates how spatial and temporal variations of the \gls{OH} skylines can
alter the data and the validity of its reduction and analysis. It is thus
mandatory to scrutinize the raw data cube to investigate the presence of
undesirable \gls{OH} skylines prior to data reduction and analysis.

\setcounter{figure}{14}

\begin{figure}[!h]
 \centerline{\includegraphics{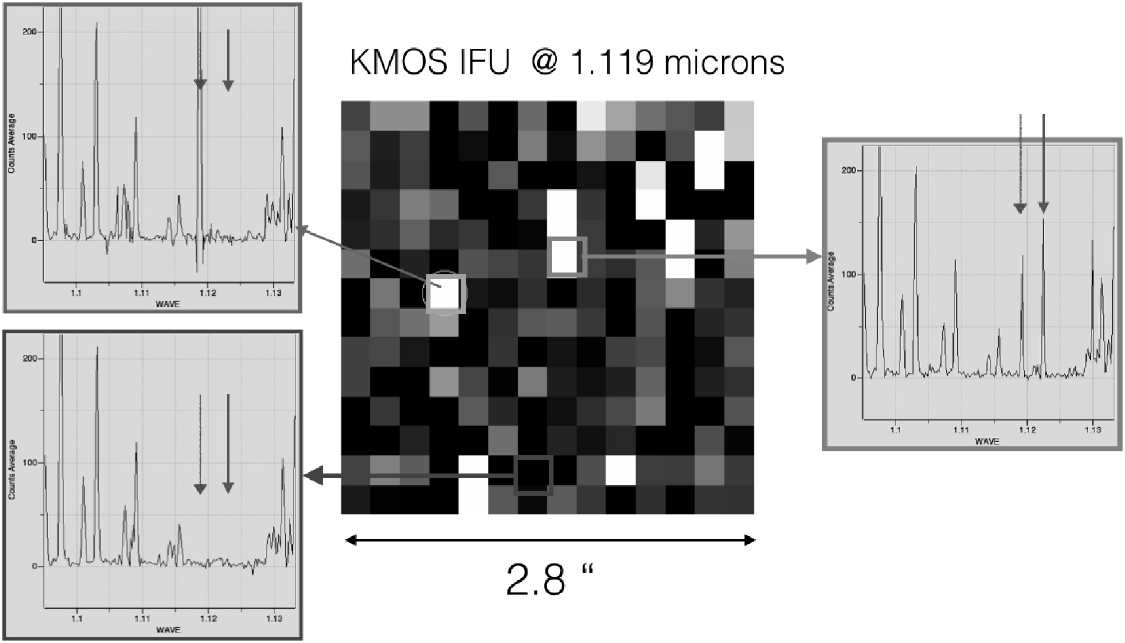}}
\caption[Variation of OH sky lines within a KMOS data cube]{Spatial
  (and sometimes temporal) variations of the \gls{OH} skylines within a
  \gls{KMOS} data cube. Although the target has been chosen to warrant
  ${\rm H}\alpha$ in a wavelength range devoid of \gls{OH} skylines, their
  variations at different positions in the \gls{KMOS} \gls{IFU} jeopardizes the
  data reduction and analysis if not accounted for. It also evidences
  the considerable temporal and spatial variations of \gls{OH} skylines,
  which justify favoring instruments with sufficiently high spectral
  resolution.\label{C4_figSkyKMOS}}
\end{figure}

\begin{figure}[h!]
 \centerline{\includegraphics{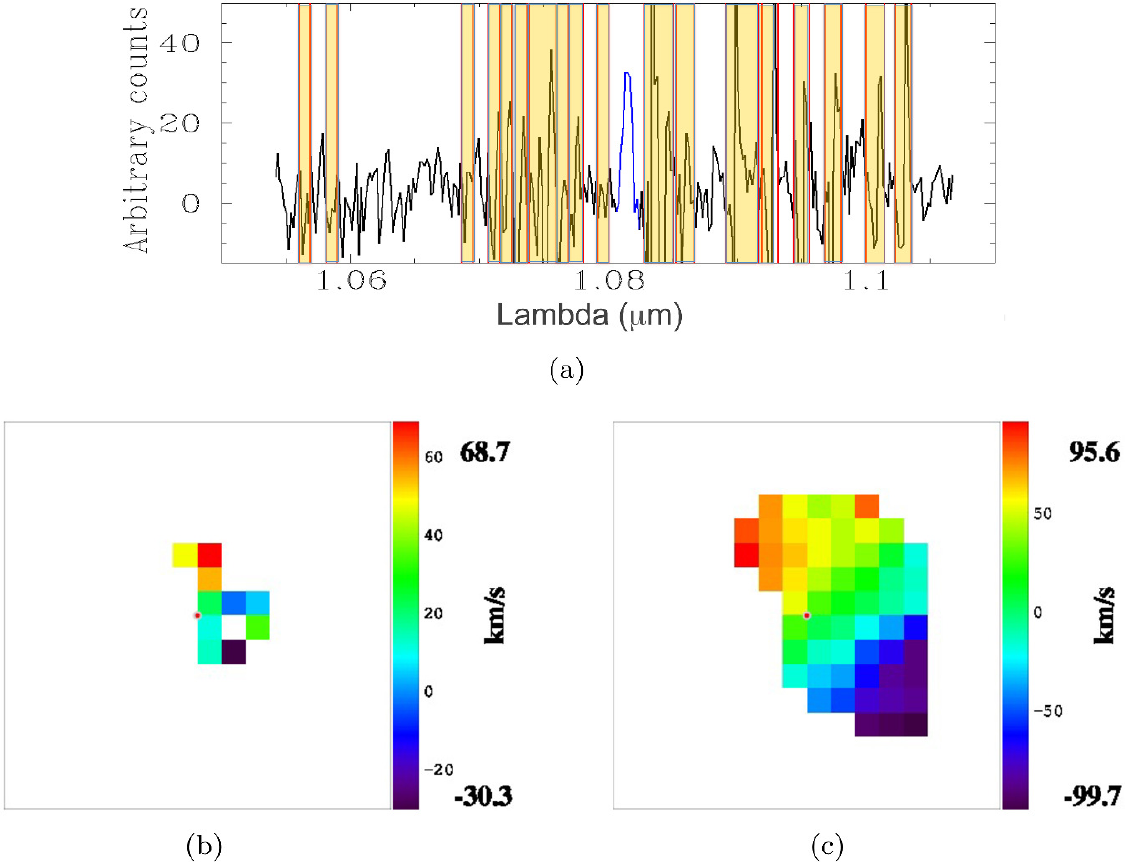}}
\caption[Optimization of the S/N for KMOS observations]{(a)
  Zoom on a spectrum provided by the sum of sky-subtracted spectra
  over the spaxels covering a z $=$ 0.649 galaxy. The \gls{S/N} in the continuum
  is estimated on a much larger spectral window as provided by \gls{KMOS}. The
  numerous spikes from sky-residuals are masked (yellow and red
  vertical lines) within $\pm$2 pixels. (b) Velocity field
  without masking residual spikes. (c) Same as (b) after
  masking the skyline residuals and applying a $3\times 3$ Gaussian filter of
  1.5 spaxel (0.3 arcsec).\label{C4_3DfigKMOS}}
\end{figure}

The following Ex. \textit{Optimizing emission line S/N in the near-IR at moderate
resolution} presents \gls{KMOS}/\gls{VLT} J band observations of a $z=0.649$ galaxy.
Figure~\ref{C4_3DfigKMOS} illustrates the considerable gain when properly removing the
skyline residuals. Combined with Fig. \ref{C4_figSkyKMOS}, it demonstrates that an R of
3000 is hardly sufficient to properly remove \gls{OH} skylines. This is because such a
spectral resolution translates into a velocity resolution of 100 km$s^{-1}$, which
corresponds to a significant fraction of the velocity gradient observed in galaxies,
especially those observed at moderate inclinations. This can be avoided using resolutions
larger than 9000 (see Fig. \ref{C4_3Dfigfitandres}).

\enlargethispage{12pt}

\setcounter{box}{3}

\begin{example}{Optimizing emission line S/N in the near-IR at moderate resolution $($KMOS$)$}
  Figure \ref{C4_3DfigKMOS}a shows the spectrum of a spaxel after data
  reduction: the H$\alpha$ line at 1.082 $\mu$m (in blue) is hardly
  recognizable within the numerous spikes of sky subtraction
  residuals. In many spaxels, the \gls{S/N} of the line is well below the
  detection threshold if the sky residuals are taken into account in
  the noise determination. Figure~\ref{C4_3DfigKMOS}b shows the
  resulting velocity field for such a nonoptimized data reduction.
  The \gls{S/N} has to be estimated in a region free of skylines, because
  sky subtraction residuals artificially enhance the noise which then
  result in underestimated \gls{S/N}s. The three-steps procedure to
  properly estimate the \gls{S/N} is as follows:

\medskip (1) Extracting the mean spectrum of the data cube and create a mask of the
  wavelength ranges affected by sky \hbox{residuals}.
Spikes (both positive
  and negative) are easy to detect with a peak detection algorithm
  because they always exhibit the same signature restricted to a single
  resolution element (see Fig.\break \ref{C4_3DfigKMOS}a): here, one negative
  spike followed by a positive spike. Flag the regions to be masked
  $\sim$ $\pm$ 2 pixels around the spikes.\footnotemark

(2) The masked data cube can then be used to properly estimate the
  \gls{S/N} in each spaxel. The data cube can be smoothed spatially
  using a Gaussian filter of width similar to the \gls{PSF} of the
  observations. Spatial smoothing can significantly increase the
  \gls{S/N} of the lines at the cost of reducing the spatial
  resolution of the observation. A Gaussian smoothing should always be
  applied in the cube after the spikes have been masked out, otherwise
  the noise would propagate over many pixels in the spectra. Here a
  Gaussian filter of 1.5 spaxel (0.3 arcsec) in a $3\times 3$ spaxel box has
  been used for an observational seeing of 0.5 arcsec \gls{FWHM}.

(3) Using the masked and smoothed data cube, the emission line of the
  object can be fitted with a Gaussian profile to measure the
  position, width, and flux. The noise is computed from the unmasked
  pixels on both sides of the line. Figure~\ref{C4_3DfigKMOS}c shows the
  final velocity field map obtained by using the spatial smoothing and
  spike rejection algorithms.
\end{example}

\setcounter{footnote}{6}

\footnotetext{This assumes that the spectral resolution is sampled by
  two pixels; the exact $\lambda$ range to be flagged depends on the spectral
  resolution, i.e., the higher is the better.}

\subsection{Measuring emission lines: methods and error budget}\label{C4_sec:emmlines}

The physical and chemical properties of the ionized gas phase in the galaxies may be
studied after measuring fluxes, positions, and profile shapes of the most prominent
emission lines, such as [\ion{O}{2}]$\lambda\lambda$3726,3729\AA, H$\beta$,
[\ion{O}{3}]$\lambda\lambda$4959,5007\AA, H$\alpha$, [\ion{N}{2}]
$\lambda\lambda$6549,6584\AA, and [\ion{S}{2}]$\lambda\lambda$6716,6731\AA. For a
preliminary analysis of a data cube, it is recommended to visualize by eye each
individual spectrum to detect possible artifacts or contamination due to sky correction
residuals or unexpected faint skylines.

\begin{figure}[!t]
 \centerline{\includegraphics{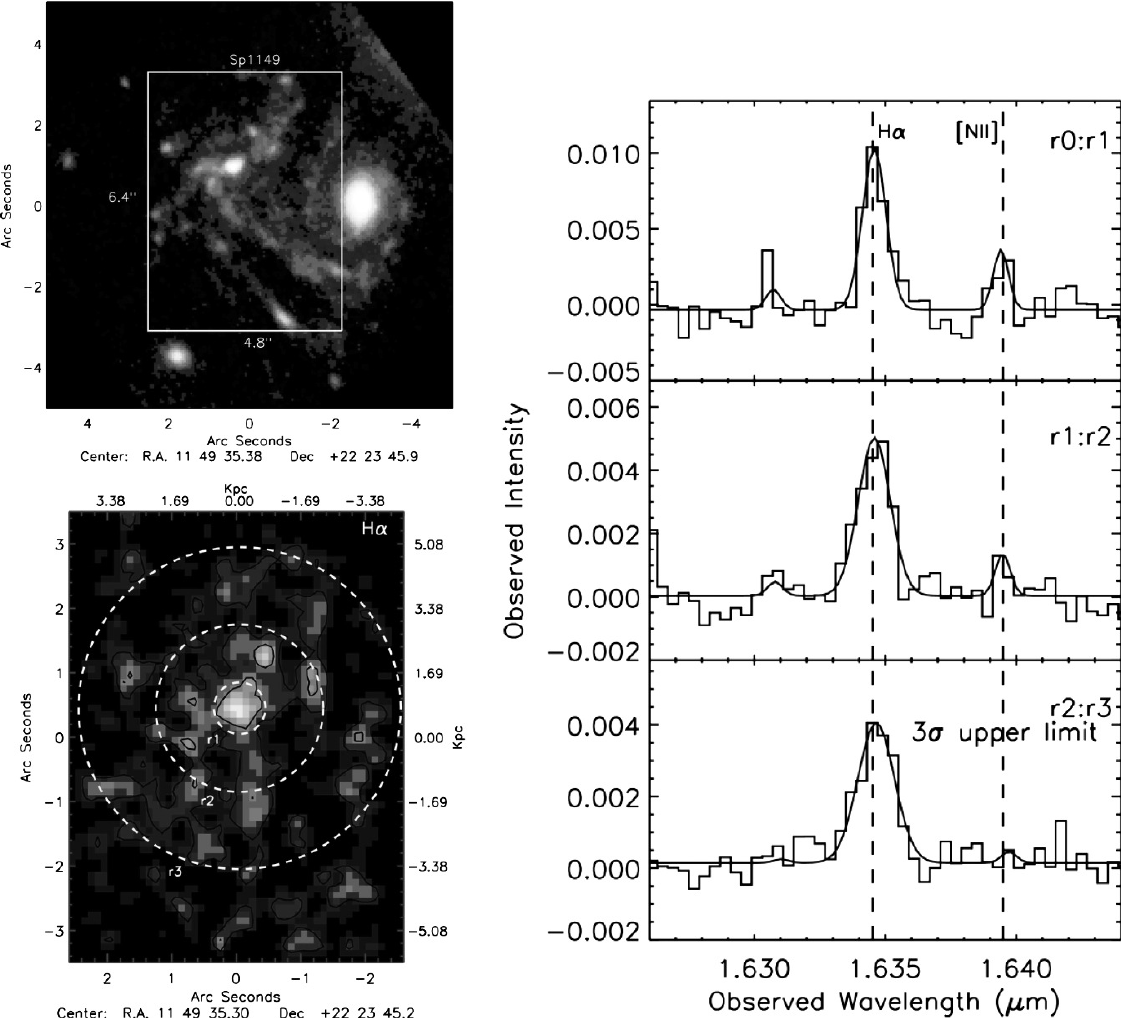}}
\caption[Example of a multiple line fit]{{\it Top-left}: \gls{HST}
  F814W and F555W band color image of a galaxy at $z=1.49$, which is
  gravitationally lensed by the MACS J1149.5+2223 cluster (from
  Yuan et al., 2011), \copyright AAS, reproduced with
  permission). The white rectangular box shows the $4.8\times 6.4$
  \gls{FoV} of the Keck/\gls{OSIRIS} \gls{IFU}. {\it Bottom-left}: the
  detected H$\alpha$ emission from the \gls{OSIRIS} datacube showing
  three rings in which the lines are searched. {\it Right}: detection
  in each ring of H$\alpha$ and [\ion{N}{2}]6548-6584 emission lines,
  allowing one to constrain the metallicity, which decreases from the
  center to the outskirts.\label{C4_HaNII}}
\end{figure}

The following example illustrates how errors can be estimated after a Gaussian fit, how
they can be optimized in some configurations, and finally how difficult it can be to
derive accurate maps of distant galaxy properties. In this case, the presence of
gravitational lensing allows robust constraints to be established on the metallicity
radial profile for a z $=$ 2.49 source (see Fig. \ref{C4_HaNII}). In general, this kind
of study can only be achieved for the brightest distant galaxies and only along an
average radial profile because of the low \gls{S/N} in the external regions of the
galaxy, and after using Voronoi binning (see Sec.~\ref{C4_spatialsmooth}).


\setcounter{box}{5}

\begin{example}{Fitting three lines together}
  Analyses of Keck/\gls{OSIRIS} 3D data have used Gaussian profiles to
  fit three emission lines ([\ion{N}{2}] $\lambda\lambda$6548-6583 and
  H$\alpha$), simultaneously (Yaun et al., 2011). The
  search for emission structures with a multiple fit permits the
  detection of even weak emissions (see Fig. \ref{C4_HaNII}). In the
  \gls{NIR}, about one out of three lines \hbox{unavoidably} lies near a strong
  \gls{OH} sky line and the procedure needs to take into account the
  correspondingly higher noise level. The line profile fitting was
  conducted using a $\chi^2$ minimization procedure, and the result
  can be checked using the fiducial ratio of the [\ion{N}{2}] lines
  ([\ion{N}{2}]6583/[\ion{N}{2}]$6548=3$). Notice that in this particular example,
  the \gls{S/N} is further enhanced by gravitational lensing.
\end{example}

Measurements of emission line fluxes are detailed in
Sec.~\ref{C3_Emission_lines_measures}. Here, we focus on how to measure the centroid and
width of the emission lines which are tracing the velocity and velocity dispersion of the
gas, respectively.

Measuring these properties requires adequate \gls{S/N} over a large
  enough number of resolution elements. The line profile \gls{S/N} can be
  defined as the mean value of the flux at a given emission as follows
  (Flores et al., 2006):
\begin{equation}
  {\rm S}/{\rm N}=\Sigma^{n}_i (S_i/(\sqrt{n}\times \sigma_{cont})),
\end{equation}
where $n$ is the number of spectral resolution elements in which the emission is
detected, $S_i$ is the emission line flux at a given voxel~$i$, and $\sigma_{cont}$ is
the noise measured on the continuum. Only spaxels with sufficient \gls{S/N} should be
considered for further analysis. For a single emission line (e.g., ${\rm H}\alpha$), a
\gls{S/N} $\ge$ 5 should be considered (see Sec.~\ref{C3_Emission_lines_measures}), while
for a doublet such as [\ion{O}{2}], the lower threshold \gls{S/N} $\ge$ 3 can be
considered acceptable (Flores et al., 2006).

At infrared wavelengths, the \gls{S/N} varies strongly as a function of wavelength
due to the numerous \gls{OH} skylines (see Fig. \ref{C4_3Dfigfitandres}). The
total noise is not only due to the sky background continuum noise but also to the
spatial residues of the correction. These originate from spatial and temporal
variations of the sky emission lines (see Fig. \ref{C4_figSkyKMOS}) and further
complicate uncertainty estimates. Figure~\ref{C4_3Dfigfitandres} further
illustrates the need for high spectral resolution in removing faint and narrow
skylines that populate spectra at $\lambda >$ 0.72 $\mu$m.


\begin{thebibliography}{99}
\bibitem[]{}
{Abraham} R.~G., {Ellis} R.~S., {Fabian} A.~C. {\it et~al.} (1999). {The
  star formation history of the hubble sequence: Spatially resolved colour
  distributions of intermediate-redshift galaxies in the hubble deep field,}
  \emph{\mnras} \textbf{303}, pp.~641--658.

\bibitem[]{}
{Cid Fernandes} R., {Mateus} A., {Sodr{\'e}} L. {\it et~al.} (2005).
  {{Semi-empirical analysis of sloan digital sky survey galaxies~--- I.
  Spectral synthesis method},} \emph{\mnras} \textbf{358}, pp. 363--378.


\bibitem[]{}
de Lapparent V., Baillard A., Bertin E. (2011). The EFIGI catalogue of 4458 nearby
galaxies with morphology. II. Statistical properties along the Hubble sequence,
\textit{A{\&}A} \textbf{532}, A75.

\bibitem[]{}
Delgado-Serrano R., Hammer F., Yang Y. B. \textit{et al.} (2010). How was the hubble
sequence 6 gyr ago? \textit{A{\&}A} \textbf{509}, A78.

\bibitem[]{}
Flores H., Hammer F., Puech M. \textit{et al.} (2006). 3D spectroscopy with VLT/GIRAFFE.
I. The true Tully Fisher relationship at z $\sim$ 0.6, {\it A{\&}A} \textbf{455},
pp.~107--118.

\bibitem[]{}
Flores H. (1999). Ph.D. thesis, Universit Paris 7 Denis Diderot.

\bibitem[]{}
Gilli R., Cimatti A., Daddi E. \textit{et~al.} (2003). Tracing the large-scale structure
in the chandra deep field south, \textit{ApJ} \textbf{592}, pp.~721--727.

\bibitem[]{}
Hammer F., Flores H., Puech M. \textit{et~al.} (2009). The hubble sequence: Just a
vestige of merger events? \textit{A{\&}A} \textbf{507}, pp. 1313--1326.

\bibitem[]{}
Hammer F., Gruel N., Thuan T. X. \textit{et~al.} (2001). Luminous compact galaxies at
intermediate redshifts: Progenitors of bulges of massive spirals? \textit{ApJ}
\textbf{550}, pp. 570--584.

\bibitem[]{}
Hogg D. W., Baldry I. K., Blanton M. R. \textit{et al.} (2002). The K correction,
\textit{ArXiv} \textit{Astrophysics e-prints}.

\bibitem[]{}
Liang Y. C., Hammer F., Flores H. \textit{et~al.} (2004). Misleading results from lowres-
olution spectroscopy: From galaxy interstellar medium chemistry to cosmic star formation
density, \textit{A{\&}A} \textbf{417}, pp.~905--918.

\bibitem[]{}
Lilly S. J., Le Fevre O., Hammer F. \textit{et~al.} (1996). The Canada--France redshift
survey: The luminosity density and star formation history of the universe to $z$
approximately~1, \textit{ApJ} \textbf{460}, p.~L1.

\bibitem[]{}
Lilly S. J., Le Fevre O., Crampton D. \textit{et~al.} (1995). The Canada--France redshift
survey. i. Introduction to the survey, photometric catalogs, and surface brightness
selection effects, \textit{ApJ} \textbf{455}, p.~50.

\bibitem[]{}
Menanteau F., Jimenez R., Matteucci F. (2001). The origin of blue cores in hubble deep
field e/s0 galaxies, \textit{ApJ} \textbf{562}, pp. L23--L27.

\bibitem[]{}
Neichel B., Hammer F., Puech M. \textit{et~al.} (2008). Images. ii. a surprisingly low
fraction of undisturbed rotating spiral disks at $z > 0 > 6$ the morpho-kinematical
relation 6 gyr ago, \textit{A{\&}A} \textbf{484}, pp. 159--172.

\bibitem[]{}
Ravikumar C. D., Puech M., Flores H. \textit{et~al.} (2007). New spectroscopic redshifts
from the cdfs and a test of the cosmological relevance of the goods-south field,
\textit{A{\&}A} \textbf{465}, pp.~1099--1108.


\bibitem[]{}
Rodrigues M., Foster C., Taylor E. \textit{et al.} (2015). Galaxy And Mass Assembly
(GAMA): Chemical evolution of star-forming galaxies from $0.07 < z < 0.34$,
\textit{A{\&}A}.

\bibitem[]{}
van den Bergh S. (2002). Ten billion years of galaxy evolution, \textit{PASP}
\textbf{114}, pp. 797--802.

\bibitem[]{}
Vanzella E., Cristiani S., Dickinson M. \textit{et~al.} (2005). The great observatories
origins deep survey. vlt/fors2 spectroscopy in the goods-south field, \textit{A{\&}A}
\textbf{434}, pp.~53--65.

\bibitem[]{}
Yang Y., Flores H., Hammer F. \textit{et~al.} (2008). IMAGES. I. Strong evolution of
galaxy kinematics since $z = 1$, \textit{A{\&}A} \textbf{477}, pp.~789--805.

\bibitem[]{}
Yuan T.-T., Kewley L. J., Swinbank A. M. \textit{et~al.} (2011). Metallicity gradient of
a lensed face-on spiral galaxy at redshift 1.49, \textit{ApJ} \textbf{732}, L14.

\bibitem[]{}
Zheng X. Z., Hammer F., Flores H. \textit{et~al.} (2004). Hst/wfpc2 morphologies and
color maps of distant luminous infrared galaxies, \textit{A{\&}A} \textbf{421},
pp.~847--862.

\bibitem[]{}
Zibetti S., Groves B. (2011). Resolved optical-infrared spectral energy distributions of
galaxies: Universal relations and their break-down on local scales, \textit{MNRAS}
\textbf{417}, pp. 812--834.

\end{thebibliography}
\end{document}